\documentclass{IEEEtran}
\usepackage{graphicx}
\usepackage{subfigure}
\usepackage[cmex10]{amsmath}
\usepackage{amssymb}
\usepackage{bm}
\usepackage{algorithm}
\usepackage{algorithmic}
\usepackage{cite}
\usepackage{setspace}
\usepackage{stfloats}
\usepackage{multirow}
\usepackage{mathrsfs}
\usepackage{color}
\begin{document}

\makeatletter
\newcommand{\rmnum}[1]{\romannumeral #1}
\newcommand{\Rmnum}[1]{\expandafter \@slowromancap \romannumeral #1@}
\makeatother

\title{Joint Optimization of Cooperative Communication and Computation in Two-Way Relay MEC Systems}

\author{Biyuan Xie, Qi Zhang, \emph{Member}, \emph{IEEE}, and Jiayin Qin

}

\markboth{}%
{Xie \MakeLowercase{\textit{et al.}}: Joint Optimization of Cooperative Communication and Computation in Two-Way Relay MEC Systems}
\maketitle

\begin{abstract}
Considering two users exchange computational results through a two-way relay equipped with a mobile-edge computing server, we investigate the joint optimization problem of cooperative communication and computation, whose objective is the total energy consumption minimization subject to the delay constraint. To derive the low-complexity optimal solution, by assuming that the transmit power of the relay is given, we theoretically derive the optimal values of computation task partition factors. Thus, the optimal solution to the total energy consumption minimization problem can be found by one-dimensional search over the transmit power of the relay. Simulation results show that our proposed scheme performs better than that where all computational tasks are allocated to the relay or users.
\end{abstract}


\IEEEpeerreviewmaketitle

\section{Introduction}

Recently, new applications, such as face recognition, natural language processing, and virtual reality, trigger the research on mobile-edge computing (MEC) \cite{LiuJ16,YouC17,WangF18,ChenX18,HuJ19}. In \cite{LiuJ16}, a power-constrained delay minimization problem in a single-user MEC system was studied. In \cite{YouC17}, resource allocation under the delay constraint for a multiuser MEC system was investigated. In \cite{WangF18}, a unified MEC-wireless power
transfer (WPT) design was proposed, by considering a wireless powered multiuser MEC system. In \cite{ChenX18}, a joint cooperative communication and computation in a
relay MEC system was put forward. In \cite{HuJ19}, for an unmanned aerial vehicle (UAV)-enabled MEC system, the joint optimization problem of UAV position, time slot
allocation, and computation task partition was solved.

In this paper, we consider a computational result exchanging system over the wireless two-way relay channel. A typical scenario is mutual identity authorization by face recognition with help of a two-way relay equipped with an MEC server. Considering cooperative communication and computation, our aim is to minimize the total energy consumption at both users and the relay, subject to the delay constraint. The formulated problem is convex and can be solved by the interior point method. Since the interior point method has high computational complexity, we propose a low-complexity optimal solution in this paper. By assuming that the transmit power of the relay is given, we theoretically derive the optimal values of computation task (CT) partition factors. With the obtained CT partition factors, we can find the optimal durations of offloading and computing. Therefore, the optimal solution to the total energy consumption minimization problem can be found by one-dimensional search over the transmit power of the relay.

\section{System Model and Problem Formulation}

Consider a computational result exchanging system over the wireless two-way relay channel. In the system, User 1 wants to share its computational results with User 2 and the results require some key parameters from User 2. User 2 also wants to share its computational results with User 1 and the results require some key parameters from User 1. We assume that the two-way relay is equipped with an MEC server and it has the global information to determine the amount of computation at both the relay and two users, respectively. We focus on a time block with duration $T$.

In the first time slot (TS) with duration $\tau_1$, User 1 offloads all of the CTs with length $L_1$ in bits to the relay over the forward relay channel $h_{1,f}$. In the second TS with duration $\tau_2$, User 2 offloads all of the CTs with length $L_2$ in bits to the relay over the forward relay channel $h_{2,f}$. The achievable data rate for offloading from User $i$, $i\in\{1,2\}$, to the relay is expressed as
\begin{equation}\label{q1}
r_{i,f}=B\log_2\left(1+P_i|h_{i,f}|^2/\sigma^2\right)
\end{equation}
where $B$ denotes the system bandwidth, $P_i$ denotes the transmit power of User $i$, and $\sigma^2$ denotes the power of the additive Gaussian noise at both the relay and users. The duration and the energy consumption for offloading from User $i$ are
\begin{equation}\label{q2}
\tau_i=L_i/r_{i,f} \mbox{ and }E_i=\tau_iP_i,
\end{equation}
respectively.

To reduce the computational delay, the two-way relay allocates CTs with length $(1-\alpha_1)L_1$ to User 2 and CTs with length $(1-\alpha_2)L_2$ to User 1 where $\alpha_i \in [0,1]$ denotes the CT partition factor for User $i$, $i\in\{1,2\}$. The remaining CTs $\alpha_1L_1+\alpha_2L_2$ are computed at the relay. Specifically, the relay employs the physical-layer network coding to simultaneously broadcast CTs with length $(1-\alpha_1)L_1$ and CTs with length $(1-\alpha_2)L_2$ to two users over the backward channels in the third TS with duration $\tau_3$. We assume that the relay employs the same coding scheme for the CTs to User 2 and User 1. Thus, we have
\begin{equation}\label{q5}
(1-\alpha_1)L_1=(1-\alpha_2)L_2.
\end{equation}
By using the local information, User 1 and User 2 are able to decode the relaying CTs. The achievable data rate at User $i$, $i\in\{1,2\}$, in the third TS is
\begin{equation}
r_{i,b} = B\log_2\left(1+P_r|h_{i,b}|^2/\sigma^2\right),\ i\in\{1,2\}
\end{equation}
where $P_r$ denotes the transmit power of the relay and $h_{i,b}$ denotes the backward relay channel from relay to User $i$. Since the relay employs the same coding scheme for the CTs to User 2 and User 1, the duration and the energy consumption of the third TS are given by
\begin{equation}\label{q7}
\tau_3=\max\{\tau_{1,b},\tau_{2,b}\}  \mbox{ and }E_3=\tau_3P_r,
\end{equation}
respectively, where
\begin{equation}\label{q8}
\tau_{1,b}=(1-\alpha_2)L_2/r_{1,b} \mbox{ and }\tau_{2,b}=(1-\alpha_1)L_1/r_{2,b}.
\end{equation}

At the fourth TS, the computing time and the energy consumption at each user are given by \cite{WangF18}
\begin{align}\label{q10}
t_{u}&=k(1-\alpha_{1})L_{1}/F_u=k(1-\alpha_{2})L_{2}/F_u,\\
C_{u}&= k(1-\alpha_{1}) L_{1}\eta_u F_u^2=k(1-\alpha_{2}) L_{2}\eta_u F_u^2,
\end{align}
respectively, where $k$ denotes the number of CPU cycles for computing one bit, $\eta_u$ denotes the effective capacitance coefficient at users, and ${F_u}$ denotes the computational speed of users \cite{WangF18}. For the relay, since both the third and fourth TS can be used for computing, we have
\begin{equation}\label{q12}
t_{r} + \tau_{3}=k(\alpha_1 L_1 + \alpha_2 L_2 )/F_r
\end{equation}
where $t_{r}$ denotes the computing time of the relay at the fourth TS and ${F_r}$ denotes the computational speed of the relay. The energy consumption at the relay is given by
\begin{equation}
C_r=k(\alpha_1 L_1 + \alpha_2 L_2)\eta_r F_r^2
\end{equation}
where $\eta_r$ denotes the effective capacitance coefficient at the relay.
After computing at the relay, the relay forwards the results to users. According to \cite{WangF18}, the result forwarding time duration is relatively small and negligible.

Considering the whole computational result exchanging process, our aim is to minimize the total energy consumption at both users and the relay, subject to the delay constraint
\begin{equation}\label{q14}
\tau_1+\tau_2+\tau_3+\tau_4\leq T
\end{equation}
where
\begin{equation}\label{q15}
\tau_4=\max\{t_{u},t_{r}\}.
\end{equation}
The system energy consumption optimization problem is \begin{subequations}\label{q20}
\begin{align}\label{q20a}
\min_{\Theta}\ & \mathcal{E} \\  
\label{q20b}\mbox{s.t.}\ & \eqref{q5},\ 0\leq\alpha_i \leq 1,\ i\in\{1,2\},\\
\label{q20c}& \eqref{q14},\  \tau_j\geq 0,\ j\in\{1,2,3,4\},   \\ 
\label{q20d}& P_i\geq 0,\ i\in\{1,2\},\ P_r\geq 0
\end{align}
\end{subequations}
where $\Theta=\{\alpha_1, \alpha_2, \tau_1, \tau_2, \tau_3, \tau_4, P_1, P_2, P_r\}$ and
\begin{equation}\label{q21}
\mathcal{E}=E_1+E_2+E_3+2C_u+C_r.
\end{equation}

\section{Optimal Solution}

Substituting \eqref{q2} into \eqref{q1}, we have
\begin{equation}\label{bq1}
P_i=\gamma_{i,f}^{-1}\left(2^{\frac{L_i}{B\tau_i}}-1\right)\geq0,\ i\in\{1,2\}
\end{equation}
where $\gamma_{i,f}=|h_{i,f}|^2/\sigma^2$. Substituting \eqref{bq1} into $E_i=\tau_iP_i$ for $i\in\{1,2\}$, we have
\begin{equation}\label{bq2}
E_i=\tau_i\gamma_{i,f}^{-1}\left(2^{\frac{L_i}{B\tau_i}}-1\right),\ i\in\{1,2\}
\end{equation}
which is a convex perspective function with respect to $\tau_i$. Similarly, we have
\begin{equation}\label{bq3}
E_3=\tau_{3}\gamma_{b}^{-1}\left(2^{\frac{(1-\alpha_{1})L_{1}}{B\tau_3}}-1\right)
\end{equation}
where $\gamma_{b}=\sigma^{-2}\min_i|h_{i,b}|^2$. Substituting \eqref{bq2} and \eqref{bq3} into problem \eqref{q20}, problem \eqref{q20} is convex and can be solved by the interior point method.

Since the interior point method has high computational complexity, we propose a low-complexity optimal solution in this paper. To proceed, we have the following proposition.

\emph{Proposition 1}: Denote the optimal solution to problem \eqref{q20} as $\Theta^\star=\{\alpha_1^\star, \alpha_2^\star, \tau_1^\star, \tau_2^\star, \tau_3^\star, \tau_4^\star, P_1^\star, P_2^\star, P_r^\star\}$. For the optimal solution to problem \eqref{q20}, we have
\begin{equation}\label{bq4}
\tau_1^\star+\tau_2^\star+\tau_3^\star+\tau_4^\star=T.
\end{equation}

\emph{Proof}: The proof is omitted for space limitation. $\hfill\blacksquare$

From Proposition 1, we replace the constraint \eqref{q14} in problem \eqref{q20} with the following constraint
\begin{equation}\label{bq5}
\tau_1+\tau_2+\tau_3+\tau_4=T.
\end{equation}
In the following, assuming that the optimal value of $P_r$ is given, we theoretically derive the optimal values of $\alpha_1$ and $\alpha_2$. In \eqref{q15}, we consider two cases, i.e., $t_{u}\geq t_{r}$ and $t_{u}\leq t_{r}$.

\subsection{The Case When $t_{u}\geq t_{r}$}

When $t_{u}\geq t_{r}$, using \eqref{q5}, we have
\begin{equation}\label{bq6}
\alpha_2=1-(1-\alpha_1)L_1/L_2.
\end{equation}
Since we assume that the optimal value of $P_r$ is given, from \eqref{q7} and \eqref{q8}, we obtain
\begin{equation}\label{bq7}
\tau_{3}=(1-\alpha_1)L_{1}/r_{b}
\end{equation}
where $r_b=\min\{r_{1,b},r_{2,b}\}$. Substituting \eqref{q10} and \eqref{q12} into $t_{u}\geq t_{r}$, we have
\begin{equation}\label{bq8}
\frac{k(1-\alpha_{1})L_{1}}{F_u} \geq \frac{k(\alpha_1 L_1 + \alpha_2 L_2 )}{F_r}-\tau_3.
\end{equation}
Substituting \eqref{bq6} and \eqref{bq7} into \eqref{bq8}, after some mathematical manipulation, we have
\begin{equation}\label{bq9}
\alpha_1\leq1-\phi
\end{equation}
where
\begin{equation}\label{bq10}
\phi=\frac{k(L_1+L_2)/F_r}{L_1\left(k/F_u+1/r_b+2k/F_r\right)}.
\end{equation}

Given the optimal value of $P_r$, the objective function of problem \eqref{q20} is rewritten as
\begin{align}\label{bq11}
\mathcal{E}&=\Psi+\xi+\zeta
\end{align}
where $\Psi=(L_1+L_2)\cdot(k\eta_u F_u^2+P_r/(2r_b))$ is a constant and
\begin{align}\label{bq12}
\xi=&E_1+E_2=\sum_{i=1}^2\tau_i\gamma_{i,f}^{-1}\left(2^{\frac{L_i}{B\tau_i}}-1\right),\\ \label{bq13}
\zeta=&\varphi(\alpha_1 L_1 + \alpha_2 L_2).
\end{align}
In \eqref{bq13}, $\varphi=k{\eta_r} {F_r}^2 - k{\eta_u} {F_u}^2 - P_r/(2r_b)$.

Since $t_{u}\geq t_{r}$, from \eqref{q10}, $\tau_4$, which is determined by $t_u$,  is a monotonically decreasing function of $\alpha_1$. Furthermore, from \eqref{bq7}, $\tau_3$ is a monotonically decreasing function of $\alpha_1$. Because of \eqref{bq5}, $\tau_1+\tau_2$ is a monotonically increasing function of $\alpha_1$. From \eqref{bq2}, $E_i$ is a monotonically decreasing function of $\tau_i$ for $i\in\{1,2\}$. Therefore, $\xi=E_1+E_2$ is a monotonically decreasing function of $\alpha_1$.

In \eqref{bq13}, if $\varphi\leq0$, $\zeta$ is a monotonically decreasing function of $\alpha_1$. Thus, $\mathcal{E}$ is a monotonically decreasing function of $\alpha_1$. The optimal value of $\alpha_1$ is $\alpha_1=1-\phi$.

If $\varphi>0$, $\zeta$ is a monotonically increasing function of $\alpha_1$ whereas $\xi$ is a monotonically decreasing function of $\alpha_1$. The optimal value of $\alpha_1$ is the solution to the following equation
\begin{align}\label{bq14}
\frac{\partial\xi}{\partial\alpha_1}+\frac{\partial\zeta}{\partial\alpha_1}=0,
\end{align}
because in \eqref{bq11}, $\Psi$ is a constant. Substituting \eqref{bq6} into \eqref{bq13}, we have $\zeta=\varphi(2\alpha_1L_1+L_2-L_1)$. Taking the partial derivatives of $\zeta$ with respect to $\alpha_1$, we have
\begin{equation}\label{bq16}
\frac{\partial\zeta}{\partial\alpha_1}=2\varphi L_1>0 \mbox{ and }\frac{\partial^2\zeta}{\partial\alpha_1^2}=0.
\end{equation}
Taking the first-order partial derivative of $\xi$ with respect to $\alpha_1$, we have
\begin{align}\label{bq17}
\frac{\partial\xi}{\partial\alpha_1} &=\frac{\partial\xi}{\partial\tau_1}\frac{\partial\tau_1}{\partial\alpha_1}+ \frac{\partial\xi}{\partial\tau_2}\frac{\partial\tau_2}{\partial\alpha_1}.
\end{align}
From \eqref{bq12}, we have
\begin{align}\label{bq18}
\frac{\partial\xi}{\partial\tau_i}=-\gamma_{i,f}^{-1}2^{\frac{L_i}{B \tau_i}}\left(\frac{L_i \ln2}{B\tau_i}-1\right)-\gamma_{i,f}^{-1}
\end{align}
for $i\in\{1,2\}$. To obtain $\partial\tau_1/\partial\alpha_1$ and $\partial\tau_2/\partial\alpha_2$, we substitute \eqref{q10} and \eqref{bq7} into \eqref{bq5} and obtain
\begin{equation}\label{bq19}
\tau_{1}+\tau_{2}=T-\omega(1-\alpha_1)L_1
\end{equation}
where $\omega=1/r_b+k/F_u$. Thus, we have
\begin{equation}\label{bq20}
\frac{\partial\tau_i}{\partial\alpha_1}=\frac{\partial\tau_i}{\partial(\tau_1+\tau_2)}\cdot\frac{\partial(\tau_1+\tau_2)}{\partial\alpha_1}=\frac{\omega L_1\cdot\partial\tau_i}{\partial(\tau_1+\tau_2)}
\end{equation}
for $i\in\{1,2\}$. In the following, we need to obtain $\frac{\partial\tau_i}{\partial(\tau_1+\tau_2)}$. Since $\tau_{1}$ and $\tau_{2}$ should be optimal solution to problem \eqref{q20}, if $\hat{\tau}=\tau_1+\tau_2$ is given, problem \eqref{q20} is reduced to
\begin{align}\label{bq21}
\min_{\tau_1,\tau_2}\ & E_1+E_2\ \
\mbox{s.t.}\ \tau_1+\tau_2=\hat{\tau}.
\end{align}
Problem \eqref{bq21} is convex, whose Lagrangian dual function is
\begin{align}
\mathcal{L}=\sum_{i=1}^2\tau_i\gamma_{i,f}^{-1}\left(2^{\frac{L_i}{B\tau_i}}-1\right)+\theta(\tau_1+\tau_2-\hat{\tau})
\end{align}
in which $\theta>0$ is an introduced Lagrangian variable. Using the Karush-Kuhn-Tucker (KKT) conditions, the optimal solution to problem \eqref{bq21} is
\begin{equation}\label{bq22}
\tau_i=\frac{L_i\ln 2}{B\left(\mathcal{W}\left(\frac{\theta\gamma_{i,f}-1}{e}\right)+1\right)},\ i\in\{1,2\}.
\end{equation}
Taking the first-order partial derivative of $\tau_i$ with respect to $\theta$, we have $\partial\tau_i/\partial\theta=\vartheta_i$ where
\begin{equation}\label{bq23}
\vartheta_i=\frac{L_1\gamma_{i,f}\ln2}{-B\left(\mathcal{W}\left(\frac{\theta\gamma_{i,f}-1}{e}\right)+1\right)^3\cdot\exp\left(\mathcal{W}\left(\frac{\theta\gamma_{i,f}-1}{e}\right)\right)}.
\end{equation}
Thus, we have
\begin{align}\label{bq24}
\frac{\partial\tau_i}{\partial(\tau_1+\tau_2)}=\frac{\partial\tau_i/\partial\theta}{\partial\tau_1/\partial\theta+\partial\tau_2/\partial\theta}=\frac{\vartheta_i}{\vartheta_1+\vartheta_2}.
\end{align}
Substituting \eqref{bq24} into \eqref{bq20}, we obtain
\begin{equation}\label{bq25}
\frac{\partial\tau_i}{\partial\alpha_1}=\chi_i=\frac{\omega L_1 \vartheta_i}{\vartheta_1+\vartheta_2}.
\end{equation}

Substituting \eqref{bq18} and \eqref{bq19} into \eqref{bq17}, we have
\begin{align}\label{cq1}
\frac{\partial\xi}{\partial\alpha_1}&=\sum_{i=1}^2 -\chi_i\gamma_{i,f}^{-1}2^{\frac{L_i}{B \tau_i}}\left(\frac{L_i \ln2}{B\tau_i}-1\right)-\chi_i\gamma_{i,f}^{-1}<0.
\end{align}
To obtain the solution to \eqref{bq14}, we take the second-order partial derivative of $\xi$ with respect to $\alpha_1$ and obtain
\begin{align}
\frac{\partial^2\xi}{\partial\alpha_1^2} &=\sum_{i=1}^2 2^{\frac{L_i}{B \tau_i}}\left(\frac{\chi_i L_i^2 (\ln2)^2}{\gamma_{i,f}B^2\tau_i^3}\right)>0.
\end{align}
This shows that $\partial\xi/\partial\alpha_1$ is a monotonically increasing function of $\alpha_1$. From \eqref{bq16}, there exists at most one solution to \eqref{bq14}. Substituting \eqref{bq16} and \eqref{cq1} into \eqref{bq14}, we have
\begin{align}\label{cq3}
\chi_2\gamma_{2,f}^{-1}2^{\frac{L_2}{B \tau_2}}\left(\frac{L_2 \ln2}{B\tau_2}-1\right)=\lambda.
\end{align}
where $\lambda=2\varphi L_1-\chi_1\gamma_{1,f}^{-1}2^{L_1/(B \tau_1)}(L_1 \ln2/(B\tau_1)-1)-\chi_1\gamma_{1,f}^{-1}-\chi_2\gamma_{2,f}^{-1}$.
We rewrite \eqref{cq3} as follows
\begin{align}\label{cq5}
\left(\frac{L_2 \ln2}{B\tau_2}-1\right)\cdot\exp\left(\frac{L_2\ln2}{B \tau_2}-1\right)=\frac{\gamma_{2,f}\lambda}{e\chi_2}.
\end{align}
From the property of Lambert $\mathcal{W}$ function, we obtain
\begin{align}\label{cq6}
\tau_2^o=B^{-1}\left[\mathcal{W}\left(\frac{\gamma_{2,f}\lambda}{e\chi_2}\right)+1\right]^{-1}L_2 \ln2.
\end{align}
Substituting \eqref{cq6} into \eqref{bq19}, we have
\begin{equation}
\alpha_1^o=1-(T-\tau_1-\tau_2^o)/(\omega L_1)
\end{equation}
where $\tau_1$ is defined in \eqref{bq22}. Therefore, if $\varphi>0$, the optimal value of $\alpha_1$ which minimizes $\mathcal{E}$ is
\begin{equation}
\alpha_1=\min\{\max\{\alpha_1^o,0\}, 1-\phi\}.
\end{equation}

\subsection{The Case When $t_{u}\leq t_{r}$}
In this case, from \eqref{bq9}, we know $\alpha_1\geq1-\phi$.
Substituting \eqref{q12} into \eqref{bq7}, we have
\begin{equation}\label{dq2}
\tau_{1}+\tau_{2}=T-k(2\alpha_1 L_1-L_1+L_2)/F_r.
\end{equation}
From \eqref{dq2}, $\tau_1+\tau_2$ is a monotonically decreasing function of $\alpha_1$. From \eqref{bq2}, $E_i$ is a monotonically decreasing function of $\tau_i$ for $i\in\{1,2\}$. Therefore, $\xi=E_1+E_2$ is a monotonically increasing function of $\alpha_1$.

In \eqref{bq13}, if $\varphi\geq0$, $\zeta$ is a monotonically increasing function of $\alpha_1$. Thus, $\mathcal{E}$ is a monotonically increasing function of $\alpha_1$. The optimal value of $\alpha_1$ is $\alpha_1=1-\phi$.

If $\varphi<0$, $\zeta$ is a monotonically decreasing function of $\alpha_1$ whereas $\xi$ is a monotonically increasing function of $\alpha_1$. Using the similar method in Subsection III-A, we have
\begin{align}\label{dq3}
\frac{\partial\xi}{\partial\alpha_1}&=\sum_{i=1}^2 -\tilde{\chi}_i\gamma_{i,f}^{-1}2^{\frac{L_i}{B \tau_i}}\left(\frac{L_i \ln2}{B\tau_i}-1\right)-\tilde{\chi}_i\gamma_{i,f}^{-1}<0
\end{align}
and $\partial^2\xi/\partial\alpha_1^2>0$, where $\tilde{\chi}_i=-\frac{2kL_1\vartheta_i}{F_r(\vartheta_1+\vartheta_2)}$ for $i\in\{1,2\}$. The solution to \eqref{bq14} is
\begin{equation}
\tilde{\alpha}_1^o=-\left(T-\tau_1-\tilde{\tau}_2^o-k(L_1-L_2)/F_r\right)/\tilde{\omega}
\end{equation}
where $\tau_1$ is defined in \eqref{bq22}, $\tilde{\omega}=-2kL_1/F_r$ and
\begin{align}
\tilde{\tau}_2^o=&B^{-1}\left[\mathcal{W}\left(\frac{\gamma_{2,f}\tilde{\lambda}}{e\tilde{\chi}_2}\right)+1\right]^{-1}L_2 \ln2
\end{align}
in which $\tilde{\lambda}=2\varphi L_1-\tilde{\chi}_1\gamma_{1,f}^{-1}2^{L_1/(B \tau_1)}(L_1 \ln2/(B\tau_1)-1)-\tilde{\chi}_1\gamma_{1,f}^{-1}-\tilde{\chi}_2\gamma_{2,f}^{-1}$.
Thus, if $\varphi<0$, the optimal value of $\alpha_1$ which minimizes $\mathcal{E}$ is
\begin{equation}
\alpha_1=\max\{\min\{\tilde{\alpha}_1^o,1\}, 1-\phi\}.
\end{equation}

\subsection{Summary of Algorithm}

To solve problem \eqref{q20}, we perform one-dimensional search over $P_r$. Given $P_r$, we obtain $r_b=\min\{r_{1,b},r_{2,b}\}$ by \eqref{q7}. When $\alpha_1=0$, from \eqref{bq5}, we have
\begin{equation}
\tau_{1}+\tau_{2}=\hat{\tau}=T-\omega\min\{L_1,L_2\}.
\end{equation}
When $\alpha_1=1$, from \eqref{dq2}, we have $\hat{\tau}=T-k(L_1+L_2)/F_r$. When $\alpha_1=1-\phi$, we have $\hat{\tau}=T-\phi\omega L_1$. In the aforementioned three conditions, the optimal $\tau_1$ and $\tau_2$, expressed in \eqref{bq22}, can be found by bisection search over $\theta$ such that $\tau_{1}+\tau_{2}=\hat{\tau}$ is satisfied.

When $\alpha_1=\alpha_1^o$, combining \eqref{cq6} and \eqref{bq22}, we have
\begin{equation}\label{dq5}
\mathcal{W}\left(\frac{\gamma_{2,f}\lambda}{e\chi_2}\right)=\mathcal{W}\left(\frac{\theta\gamma_{2,f}-1}{e}\right).
\end{equation}
The right-hand side of \eqref{dq5} is a monotonically increasing function of $\theta$. For the left-hand side of \eqref{dq5}, we have
\begin{equation}\label{dq6}
\frac{\partial \lambda}{\partial \tau_1}=\chi_1\gamma_{1,f}^{-1}\tau_1^{-1}2^{\frac{L_1}{B \tau_1}}\left(\frac{L_1 \ln2}{B\tau_1}\right)^2>0.
\end{equation}
Furthermore, from \eqref{bq22}, $\tau_1$ is a monotonically decreasing function of $\theta$. Thus, the left-hand side of \eqref{dq5} is a monotonically decreasing function of $\theta$. Solving \eqref{dq5} by the bisection search over $\theta$, we can find the optimal $\tau_1$ and $\tau_2$.

Similarly, when $\alpha_1=\tilde{\alpha}_1^o$, solving the following equation
\begin{equation}\label{dq7}
\mathcal{W}\left(\frac{\gamma_{2,f}\tilde{\lambda}}{e\tilde{\chi}_2}\right)=\mathcal{W}\left(\frac{\theta\gamma_{2,f}-1}{e}\right),
\end{equation}
we can find the optimal $\tau_1$ and $\tau_2$.

After obtaining $\tau_1$ and $\tau_2$, we obtain $\tau_3$ using \eqref{bq7} and then $\tau_4$ using \eqref{bq5}. The optimal transmit power of User 1 and User 2, i.e., $P_1$ and $P_2$, can be found by \eqref{bq1}.

Comparing the five conditions of $\alpha_1$, we obtain the optimal system energy consumption $\mathcal{E}$ given $P_r$. By performing one-dimensional search over $P_r$, we obtain the optimal solution to problem \eqref{q20}.

\section{Simulation Results}

In the simulations, both the forward and backward channels are modeled as independent Rayleigh fading with average power loss $10^{-6}$. The power of the additive Gaussian noise at both the relay and users is $10^{-9}$ W \cite{WangF18}. The system bandwidth is $1$ MHz. The length of CTs is $L_1=L_2=1.8\times 10^5$ bits. The number of CPU cycles for computing one bit is $k=10^3$ cycles/bit. The effective capacitance coefficients are $\eta_u=\eta_r=10^{-28}$ Joule/(cycles$\cdot$Hz$^2$) \cite{WangF18}. The computational speed of users is $F_u=0.3$ GHz and that of the relay is $F_r=0.6$ GHz. In Fig. 1, we compare the total energy consumption of the proposed scheme with the ``Relay Computing" and ``Local Computing" schemes where the ``Relay Computing" scheme means that all CTs are allocated to the relay, i.e., $\alpha_1=\alpha_2=1$, and the ``Local Computing" scheme means that all CTs are allocated to users, i.e., $\alpha_1=\alpha_2=0$. From Fig. 1, it is found that our proposed scheme significantly outperforms the ``Relay Computing" and ``Local Computing" schemes.

\begin{figure}
\centering{\includegraphics[width=3.25in]{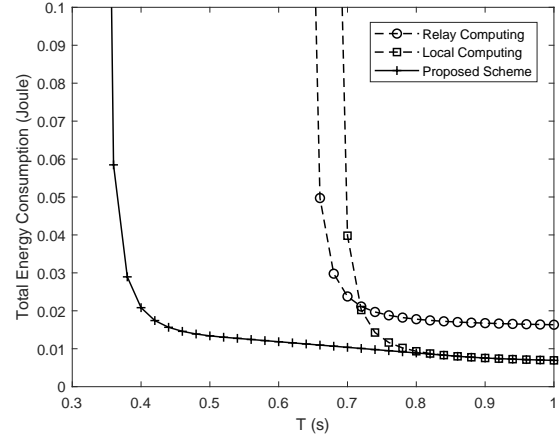}}
\caption{Total energy consumption versus the delay constraint $T$; performance comparison of different schemes.}
\end{figure}

\section{Conclusion}

In this paper, we have proposed a low-complexity algorithm to solve the total energy consumption minimization problem for a computational result exchanging system. Simulation results illustrate that our proposed scheme performs better than that where all CTs are allocated to the relay or users.

\end{document}